\documentclass[final,sort&compress,5p,reprint,preprintnumbers]{elsarticle}

\usepackage{graphicx}
\usepackage{dcolumn}
\usepackage{bm}
\usepackage{amsmath}
\usepackage{multirow}

\newcommand{\mean}[1]{\mbox{$\langle{#1}\rangle$}}

\usepackage[colorlinks=true]{hyperref}
\usepackage{amsmath}
\usepackage{amssymb}
\usepackage{array,booktabs}
\usepackage{graphicx}
\usepackage{subcaption}

\newcommand{\SPIFFE}{{\sc spiffe}}
\begin{document}

\begin{frontmatter}

\title{Initial Beam Dynamics Simulations of a High-Average-Current Field-Emission Electron Source in a Superconducting RadioFrequency Gun }
\author[niu]{O. Mohsen}
\author[apc]{I. Gonin}
\author[apc]{R. Kephart}
\author[apc]{T. Khabiboulline}
\author[niu,apc]{P. Piot}
\author[apc]{N. Solyak}
\author[apc]{J. C. Thangaraj}
\author[apc]{V. Yakovlev }
\address[niu]{Department of Physics and Northern Illinois Center for Accelerator \& Detector Development, \\
Northern Illinois University, DeKalb, IL  60115, USA}
\address[apc]{Fermi National Accelerator Laboratory, Batavia, IL  60510, USA}

\begin{abstract}
High-power electron beams are sought-after tools in support to a wide array of societal applications. This paper investigates the production of high-power electron beams by combining a high-current field-emission electron source to a superconducting radio-frequency (SRF) cavity.  We especially carry out beam-dynamics simulations that demonstrate the viability of the scheme to form $\sim 300$~kW average-power electron beam using a 1+1/2-cell SRF gun.
\end{abstract}


\end{frontmatter}


\section{Introduction}
Electron accelerators are finding an increasing number of applications in our society. Many of these applications require high average beam power produced at low cost and within compact footprints. A superconducting radiofrequency (SRF) cavity combined with a high-current electron source offer a promising path to the generation of multi~MeV Amp\`ere-class electron beams within meter-scale  footprints~\cite{Kephart}.

Thermionic cathodes are well established as a robust technology but their operation in RF guns is prone to back bombardments and beam losses~\cite{kii2001experiment}. The mitigation of back bombardment is challenging and its total suppression has been elusive so far.  Another complication associated to thermionic cathodes is their high-temperature operation which poses engineering challenge when combined with the cryogenic temperatures necessary for SRF cavities operation~\cite{lewellen2005field}. Photoemission sources are capable of producing bright electron beams but require a laser system to trigger the emission which increases the complexity, robustness and cost of the source.

\begin{figure}[bbb!!!!]
\centering\includegraphics[width=0.48\textwidth]{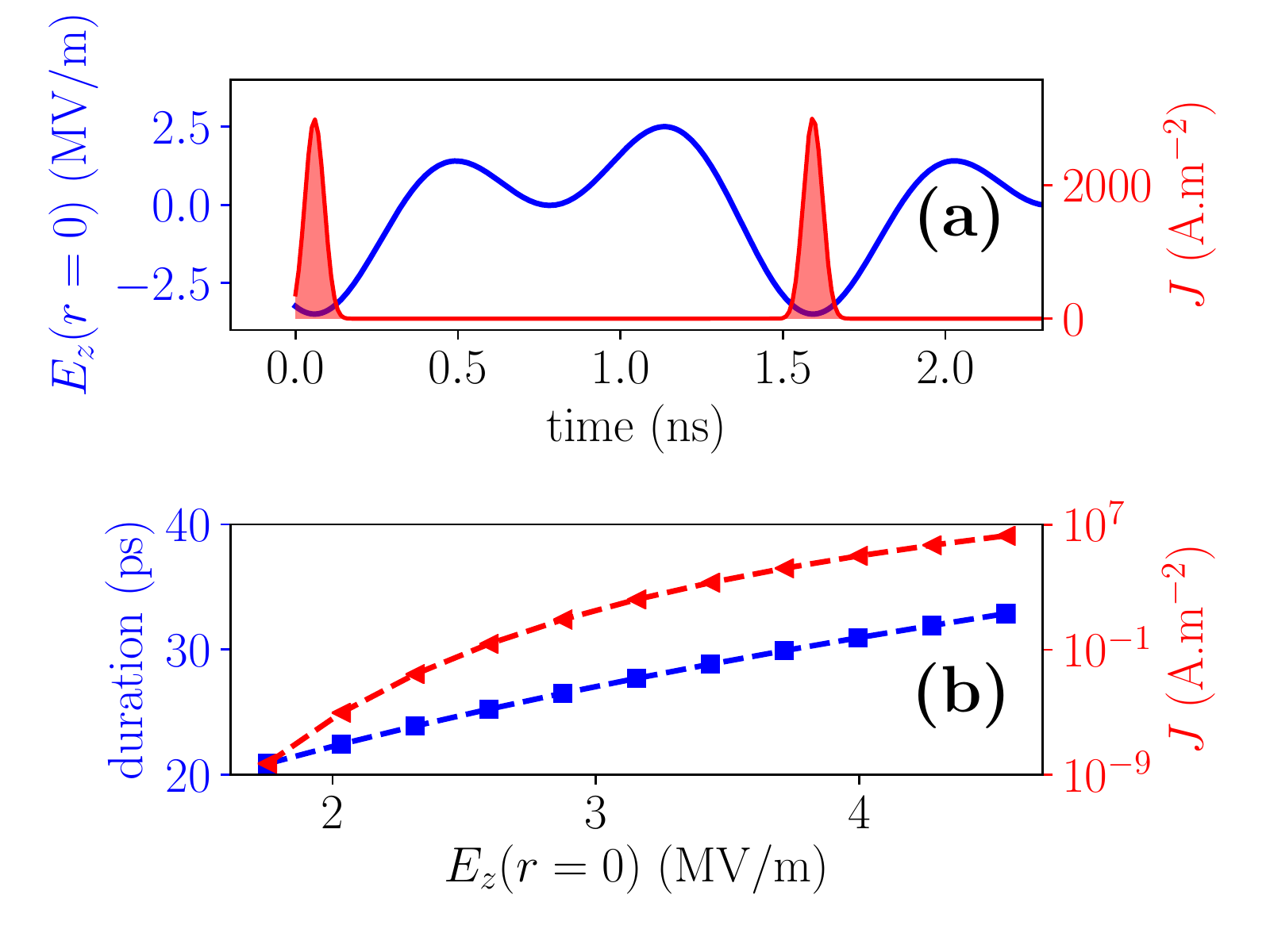}
\caption{Axial electric field applied on the cathode (blue trace) and the emitted current density (red shaded curve) (a). Evolution of the current density peak value (squares) and rms electron pulse duration (triangles) as function of the applied peak field (b). The peak-field configuration and associated parameters are detailed in Sec.~\ref{sec:concept}. Note that these are ideal numbers calculated at the cathode surface without accounting for possible saturation effects of the emission process.}
\label{fig:FN}
\end{figure}
Field-emission (FE) sources offer an appealing alternative due to the simplified setup (no auxiliary system, no heat load). Likewise, FE occurs when the electric field pulls and accelerates the electrons {\em away} from the emitting surface thereby intrinsically gating the emission process. Consequently, FE is expected to be less prone  to back bombardment and beam loss during acceleration in the SRF cavities. FE has gained much interest over the last decade owing to new development in field-emission displays and vacuum electronic devices. Additionally, FE has also been extensively investigated as electron sources for bright~\cite{Jarvis} or high-current~\cite{piotAPL} electron beams. The theory behind FE is well established and the current density is related to the applied electric field $E_z(t)$ via the Fowler-Nordheim's law~\cite{FN1,FN2}
\begin{align}\label{eq:FN}
j(t) & = A(\Phi) [\beta_e E_z(t)]^2 e^\frac{-B(\Phi)}{\beta_e E_z(t)},
\end{align}
where the parameters $A(\Phi)$ and $B(\Phi)$ depend on the work function $\Phi$  of the material and $\beta_e$ is the field enhancement factor.  When the field is time dependent the current density is modulated and the bunched electron beams are emitted; see Fig.~\ref{fig:FN}(a). The latter Figure also displays the bunch duration and current density evolution as a function of the applied peak electric field.

\section{Concept \label{sec:concept}}
\begin{figure}[hhh!!!]
\centering\includegraphics[width=0.485\textwidth]{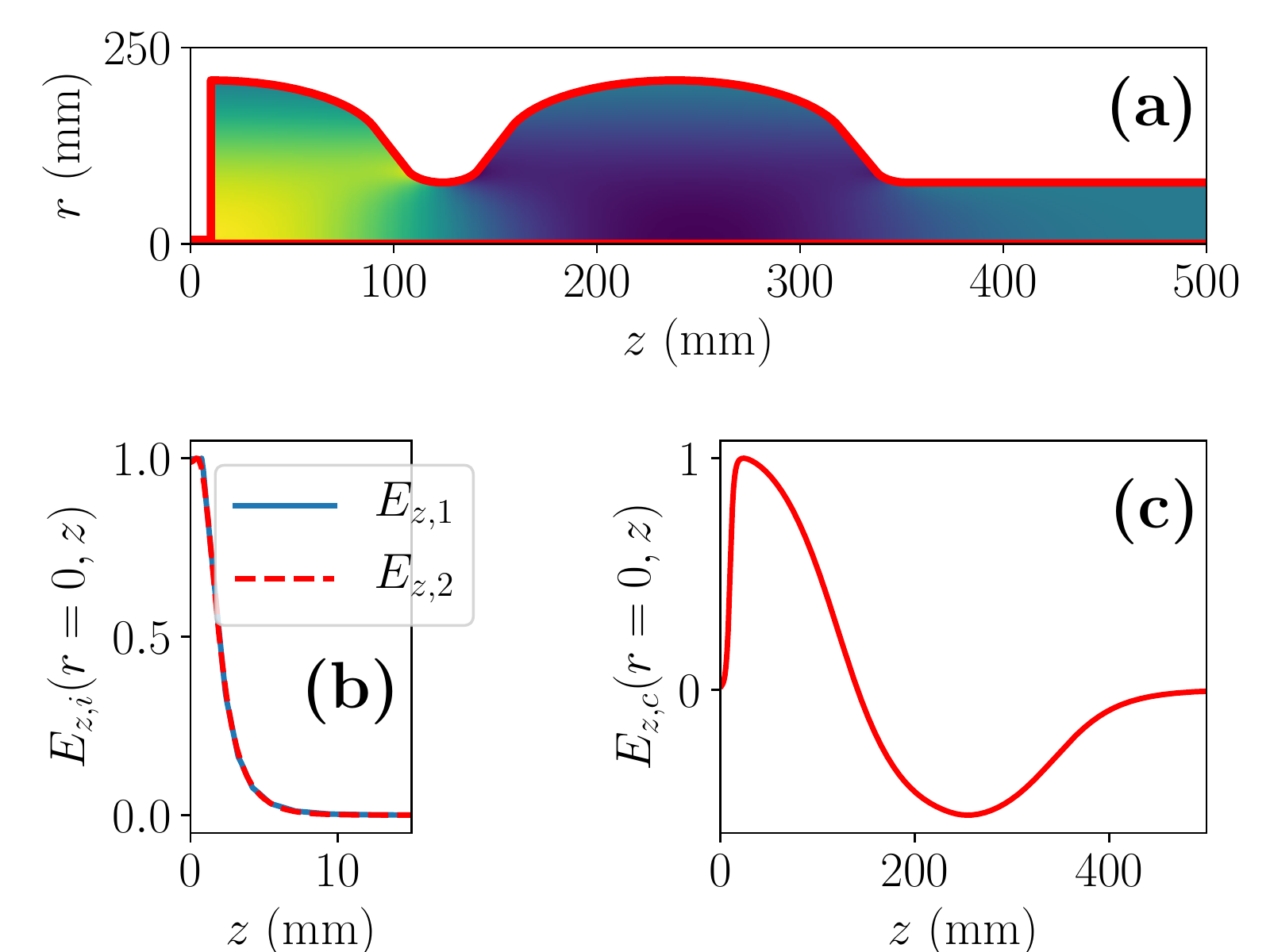}
\caption{Geometry of the SRF cavity (red trace) with field distribution $E_z(r,z)$ (a), axial fundamental $E_{z,1} (r=0,z)$ and harmonic $E_{z,2} (r=0,z)$ electric fields in the vicinity of the cathode surface and axial field  $E_{z,c} (r=0,z)$ along the 1+1/2-cell cavity (c). All the field are normalized to their peak amplitudes.  }
\label{fig:cav}
\end{figure}

Fermilab is developing technologies to support the development of compact SRF MW-class electron accelerator. A critical component toward toward the realization of small footprint is the use of conduction cooling employing cryocooler~\cite{Kephart} thereby circumventing the use of a large cryogenic infrastructure. A contemplated approach toward the development of MW-class electron source explored at Fermilab  uses a high-current electron source coupled to a 650-MHz SRF cavity. In order to reach the needed energy $\sim10$-MeV energy a $4+1/2$-cell SRF cavity is envisioned. In a first stage a $1+1/2$-cell prototype implementing conduction cooling will be tested and the produced electron beam characterized. In its nominal configuration the SRF cavity is foreseen to incorporate a thermionic electron source. However, a FE-source option is also under consideration pending the associated technological risk can be mitigated. In this paper, we use a simple cavity model for a 1+1/2-cell SRF gun with its geometry obtained from  a frequency scaling of the 1.3~GHz TESLA-cavity geometry~\cite{tesla} such to resonate at 650~MHz; see Fig.~\ref{fig:cav}(a).
The model was simulated with the {\sc superfish} software~\cite{sfish}. One peculiarity of the SRF-gun design regards the addition of an aperture on its backplate to allow for the high-current electron source to be recessed. This setup enables the cathode to be located a few mm behind the SRF-cavity backplate and subjected to a tailored electromagnetic (and possibly electrostatic) field as discussed in Ref.~\cite{lewellen2005field,edelen}. Such a configuration allows for the independent control of the emission and acceleration processes, it also avoids direct exposure of the cathode to in the SRF-cavity field thereby mitigating risks associated to vacuum contamination. In our investigation, we consider the cathode to be located at the extremity of a small RF gap with net on-axis axial electromagnetic field given by the superimposition of the SRF cavity field with a local fundamental and a second-harmonic fields
\begin{align}
E_z(z,t) & = E_{z,c}(z)\sin (2\pi f t) + E_{z,1}(z)\sin (2\pi f t + \varphi_1)  \nonumber \\
 & ~~~~~~~~ + E_{z,2}(z)\sin (4\pi f t + \varphi_2),
\end{align}
where $f=650$~MHz is the fundamental frequency, $E_{z,c}$ is the penetrating field from the SRF cavity while $E_{z,i}$ and $\varphi_i$ are respectively the field amplitudes and phases of the fundamental ($i=1$) and harmonic ($i=2$) fields produced by a bimodal cavity~\cite{edelen}. The axial coordinate  $z$ is referenced w.r.t. the cathode surface. As a first iteration the bimodal cavity designed in support of the thermionic-source option was used and its field accordingly scaled. Figure~\ref{fig:cav}(b) displays the normalized axial fundamental and second-harmonic fields supported by the bi-modal cavity as modeled with {\sc smason}~\cite{smason}. In order to mimic the boundary conditions imposed by the RF gap on SRF-cavity field, a cylindrical pipe was added at the extremity of the cavity $z \in[0, 10]$~mm. The field in the 650~MHz cavity was also simulated with the finite-difference time-domain (FDTD) axially-symmetric program \SPIFFE~\cite{borland1992summary} to provide two-dimensional $(r,z)$  field maps for the beam-dynamics simulations.
\section{Beam-dynamics simulations}
\subsection{Methods}
The performances of the FE source were explored via numerical simulations carried out with \SPIFFE~which incorporates a particle-in-cell (PIC) algorithm. The field supported by the bimodal cavity were modeled from their axial spatial profile $E_i(z)$ and the associated transverse electromagnetic fields are obtained within \SPIFFE~using the paraxial approximation. However, given the possible large transverse orbit excursion in the SRF cavity, the SRF-cavity field were described by the time-dependent 2-D field map obtained using \SPIFFE~ ran in the electromagnetic mode. The time-dependent 2-D map was then shifted to ensure sure the field of cavity is phased to allow emission for at the times $t_m=m/f=1.46 \times m \times 10^{-9}$~sec where $m$ is an integer; see Fig.~\ref{fig:FN}.

The FE process based on Eq,~\ref{eq:FN} is implemented in \SPIFFE. The cathode is located at $z=0$ and its outer radius was taken to be 3 mm. The work function was taken to be $\Phi=4.5$~eV (carbon-based field emitter) and a field enhancement factor $\beta_e=700$ was selected consistent with, e.g., values observed in carbon-nanotube field emitters~\cite{piotAPL}. We note that lower field-enhancement factors could be employed at the expense of increasing the peak field on the cathode. Such an optimization will be performed together with optimizing possible solutions for the bimodal-cavity design~\cite{jay}. In order to reduce the simulation time, the macro-particles representing the bunch consist of 10000 electrons. Owing to the self-phasing of the field-emission process, our simulation use the phase of the SRF cavity as a reference and the phases of the fields supported by the bimodal cavity are referenced w.r.t. the SRF-cavity phase.  The phases and amplitude of the fundamental and harmonic fields in the RF gap were all set to be equal; specifically  $\varphi_1 = \varphi_2 \equiv \phi_{gun}$ and $E_{z,1}(0)=E_{z,2} (0)=2$~MV/m.
\subsection{Results}
\begin{figure}[h!]
\centering\includegraphics[width=0.49\textwidth]{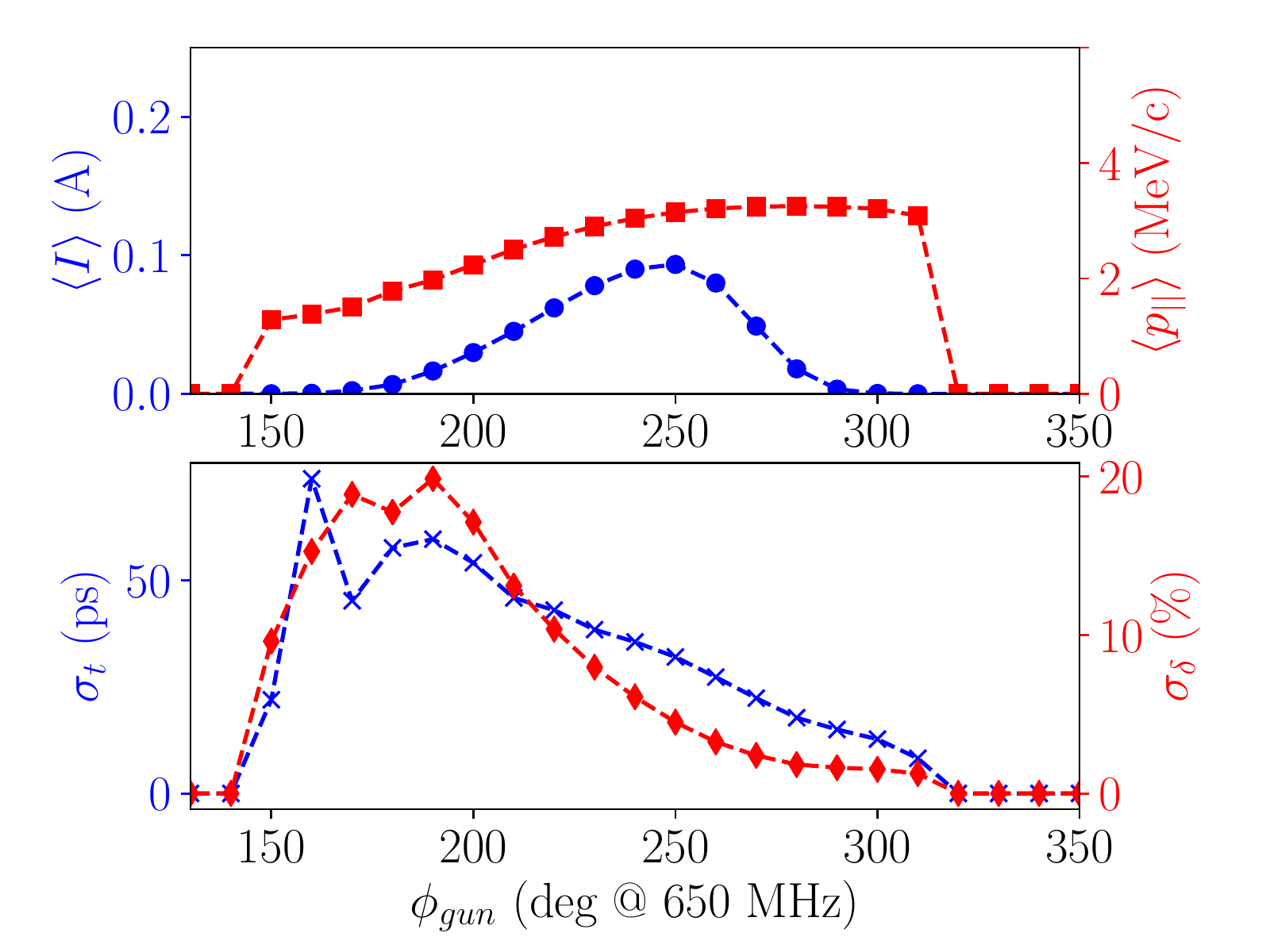}
\caption{Beam parameters verses phase of the gun. Top plot: average beam current (Blue) and average longitudinal momentum (red). Bottom plot: RMS bunch duration (blue) and fractional momentum spread (red)}
\label{Figure_2}
\end{figure}

A critical parameters to the beam dynamics is the selected emission ``launch" phase $\phi_{gun}$ w.r.t. the SRF cavity field. Figure~\ref{Figure_2} displays the evolution of the final beam energy, and other longitudinal-phase-space parameters as a function of  $\phi_{gun}$. The results indicate that a maximal average current of $\mean{I}\simeq 93$~mA can be attained with a mean momentum of $\mean{p_z}\simeq 3.15$~MeV/c for a launch phase of $\phi_{gun}=250^{\circ}$.  Figure ~\ref{fig:FN}(a) gives the corresponding axial field  on the cathode as a function of time with the emitted current density.  The corresponding bunch duration and fractional momentum spread are respectively $\sigma_t\simeq 35$~ps and $\sigma_{\delta}\simeq 5$\%.
This set of beam parameters correspond to an average-power of $\mean{P}\simeq 292$~kW. For this maximum-current launch phase our simulation did not result in any back bombardment nor particle losses which, given the number of macroparticle used in our simulation $N_m \simeq 90000$, sets an upper limit for the relative charge loss of $Q_{loss}/Q < 1/N_m \simeq 1.11\times 10^{-5}$. The latter value provide an estimate on the upper limit for the average power loss of $\mean{P_{loss}}\simeq Q_{loss}/Q \mean{P} < 3.2$~W  within the cryogenic budget.  It should be stressed that this value is an upper limit which is limited by the reduced number of macroparticle used in our simulation. Carrying out large-scale simulations, once the design is refined, will provide a better resolution on the power loss.

\begin{figure}[hhh!!!]
\centering\includegraphics[width=0.49\textwidth]{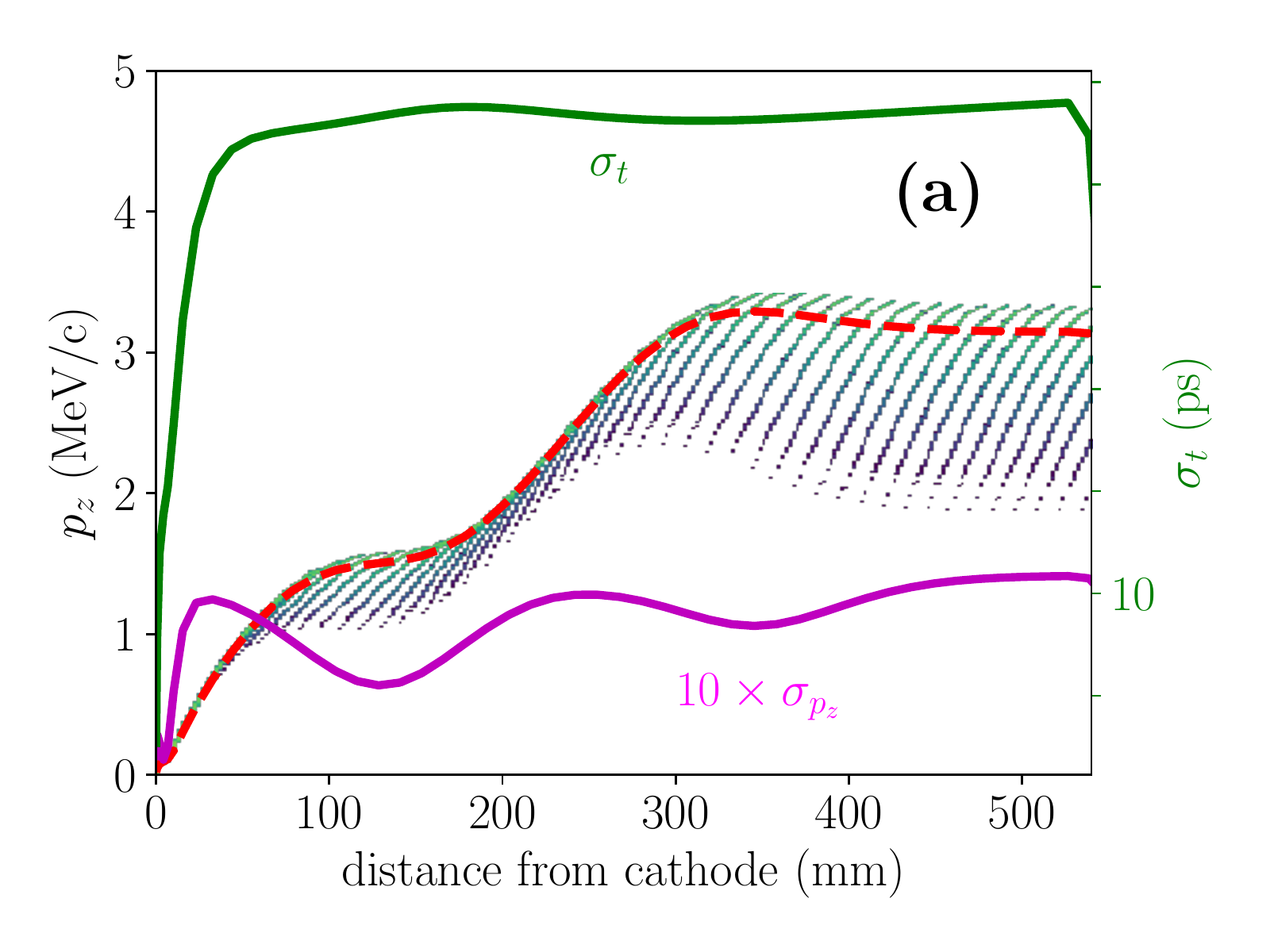}
\centering\includegraphics[width=0.49\textwidth]{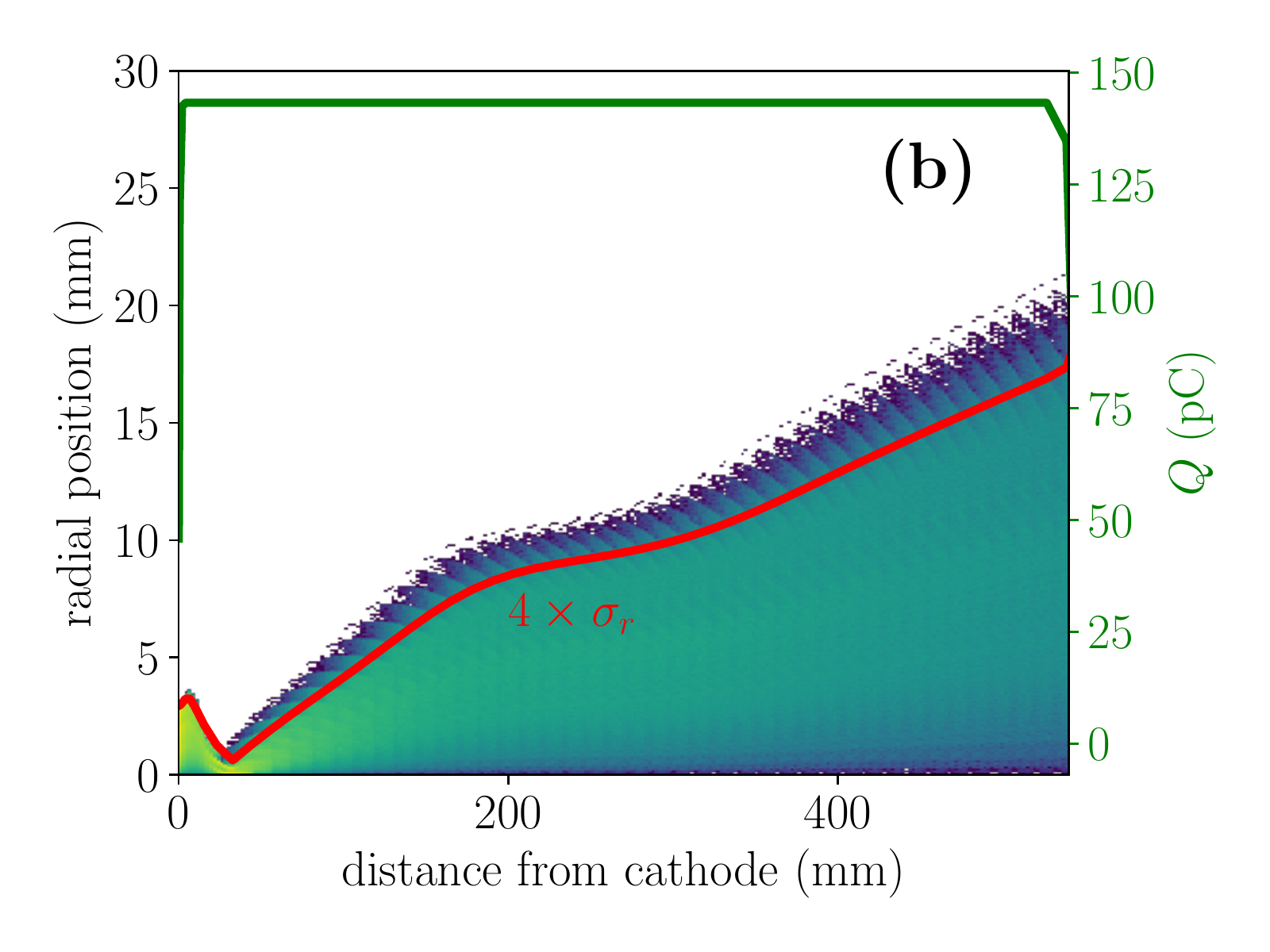}
\caption{Evolution of the longitudinal momentum (top) and radial (lower) distribution along the SRF cavity (color density histograms). The traces on the top plot represent the RMS bunch duration (green), the mean longitudinal momentum (dash red) and 10 times the RMS momentum (magenta).  The traces on the lower plot respectively give 4 times the RMS radius (red) and the bunch charge (green). }
\label{Figure_5}
\end{figure}

The corresponding evolutions of the longitudinal momentum and radial distributions during acceleration in the SRF cavity appear in Fig.~\ref{Figure_5} along with associated RMS quantities. The bunch duration reaches its final value within the first half cell of the SRF cavity while the fractional energy spread continuously increases as the longitudinal-phase-space nonlinearity impressed by the RF waveform accumulates during the acceleration process. \\  Given the absence of focusing the 4$\times \sigma_r$ beam radius continuously increases to its final value of $4 \times \sigma_r \simeq 17$~mm at the exit of the SRF-cavity. While such a value is well within the aperture sets by the beamline vacuum-pipe radius, schemes to mitigate the beam size growth will be explored (these include modifying the 1/2-cell of the SRF cavity to introduce some RF focusing).

\begin{figure}
\centering
\begin{subfigure}[b]{1\linewidth}
\includegraphics[width=43mm]{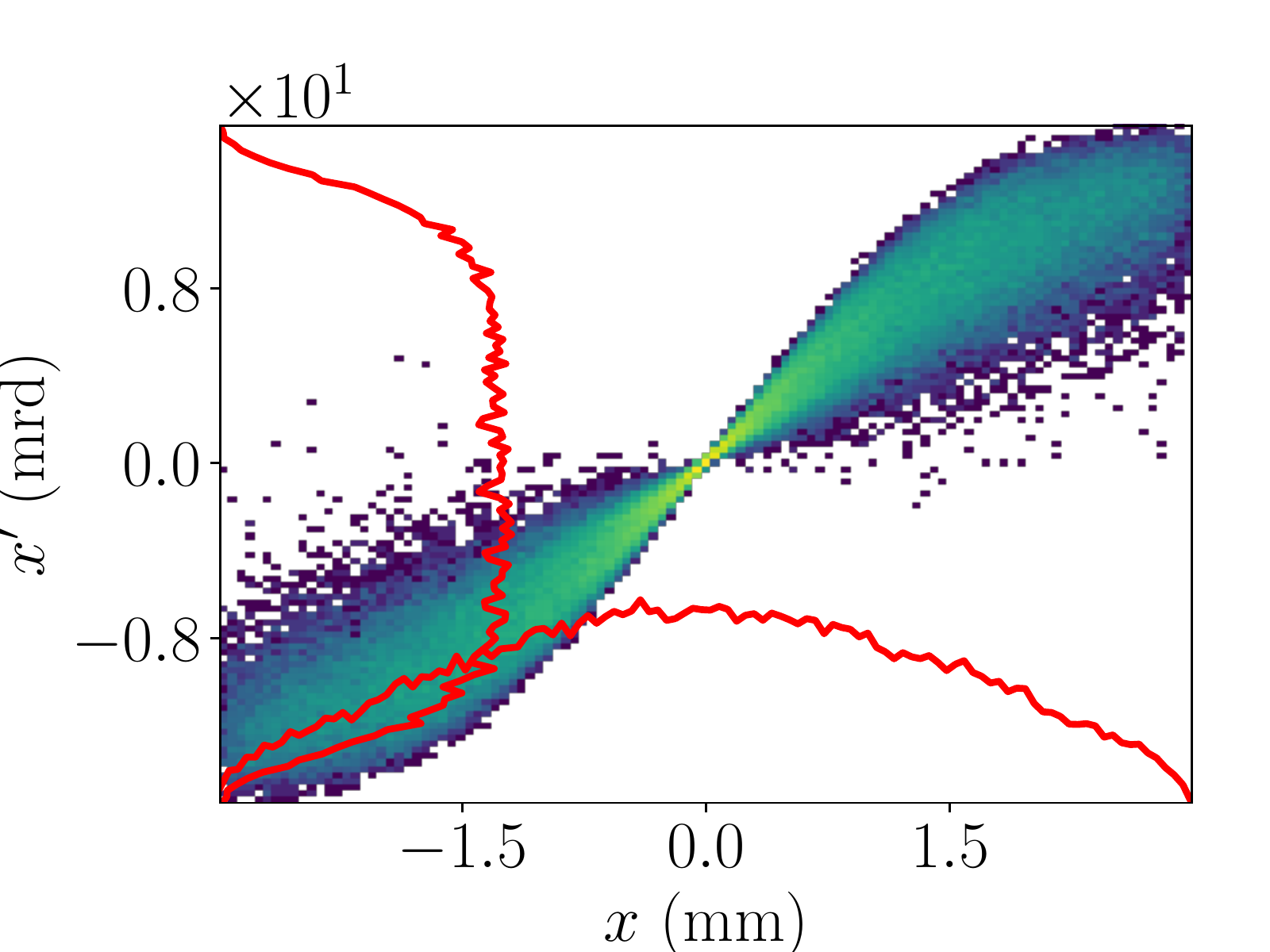}
\includegraphics[width=43mm]{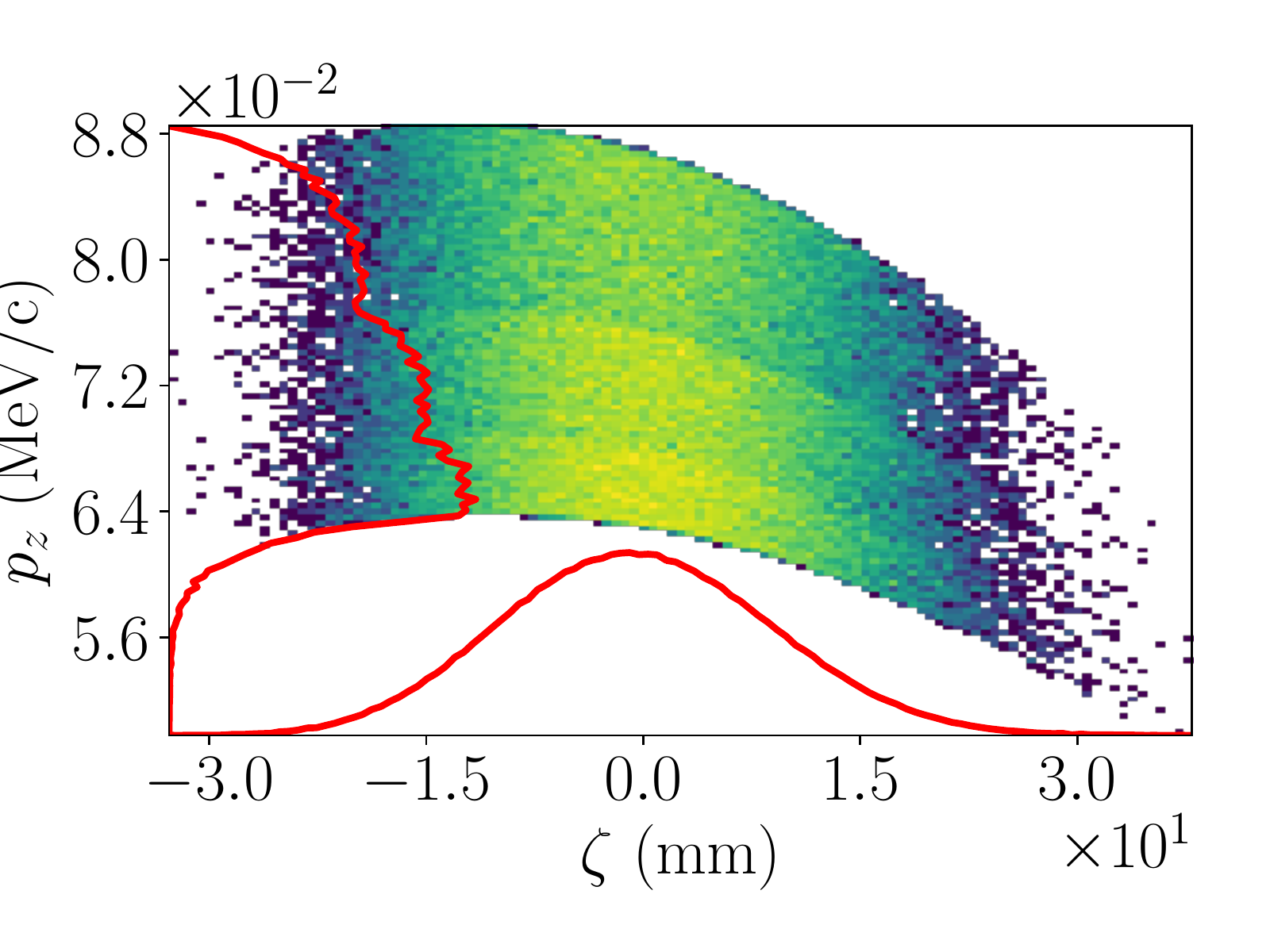}
\caption{$z= 1$~mm }
\label{subfig:cat}
\end{subfigure}
\begin{subfigure}[b]{1\linewidth}
\includegraphics[width=43mm]{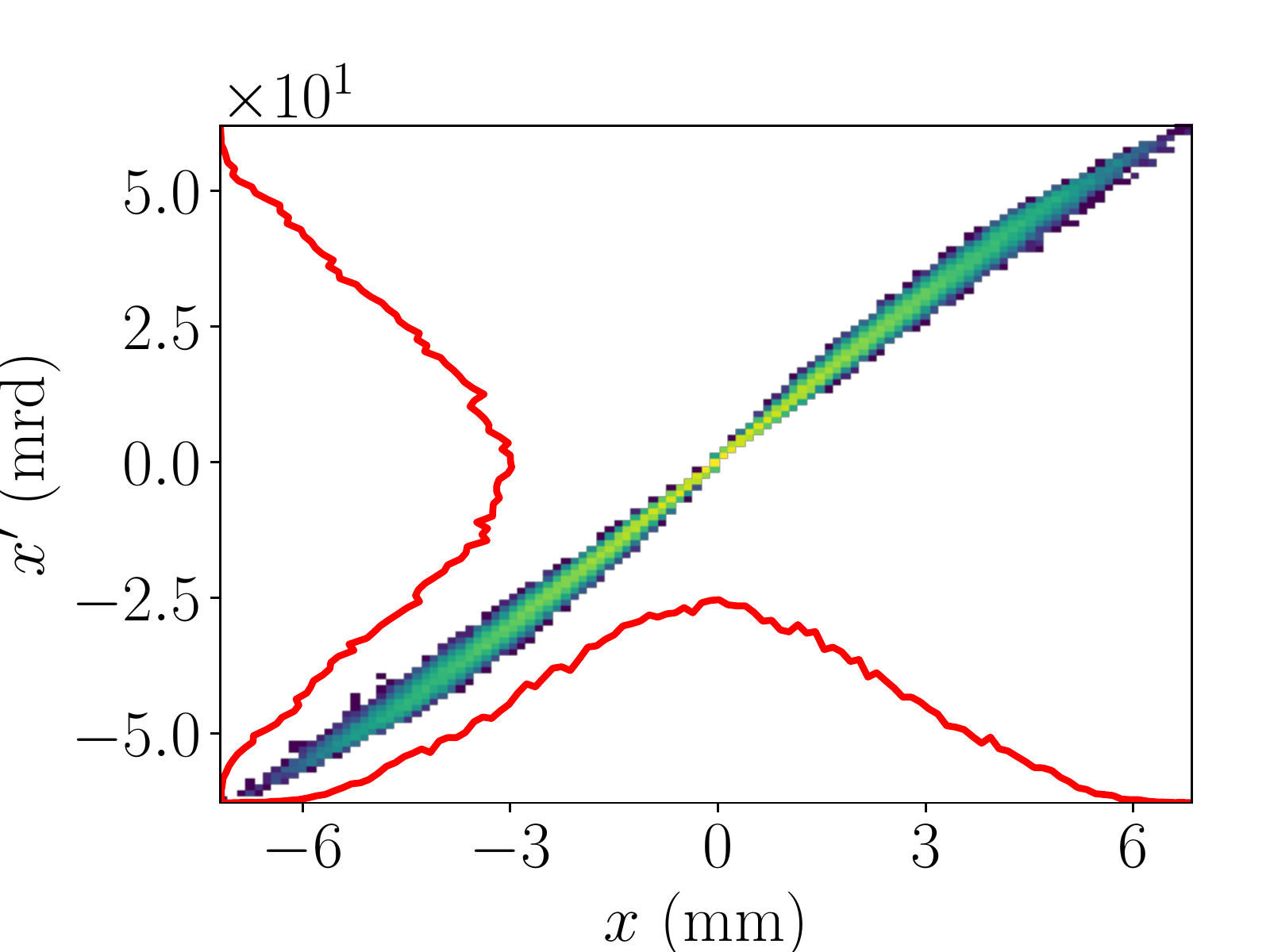}
\includegraphics[width=43mm]{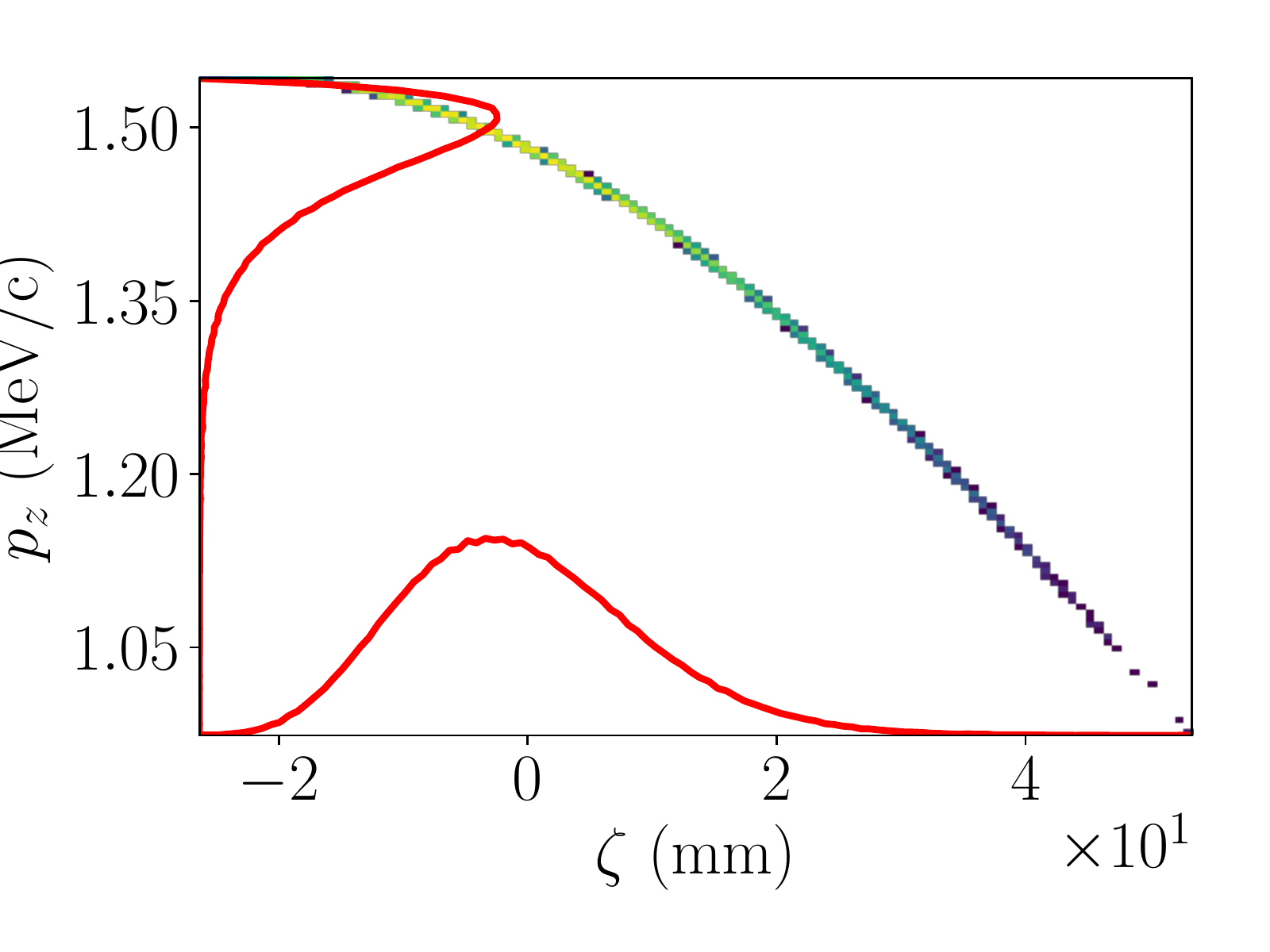}
\caption{$z=124$~mm}\label{fig:DM}
\end{subfigure}
\begin{subfigure}[b]{1\linewidth}
\includegraphics[width=43mm]{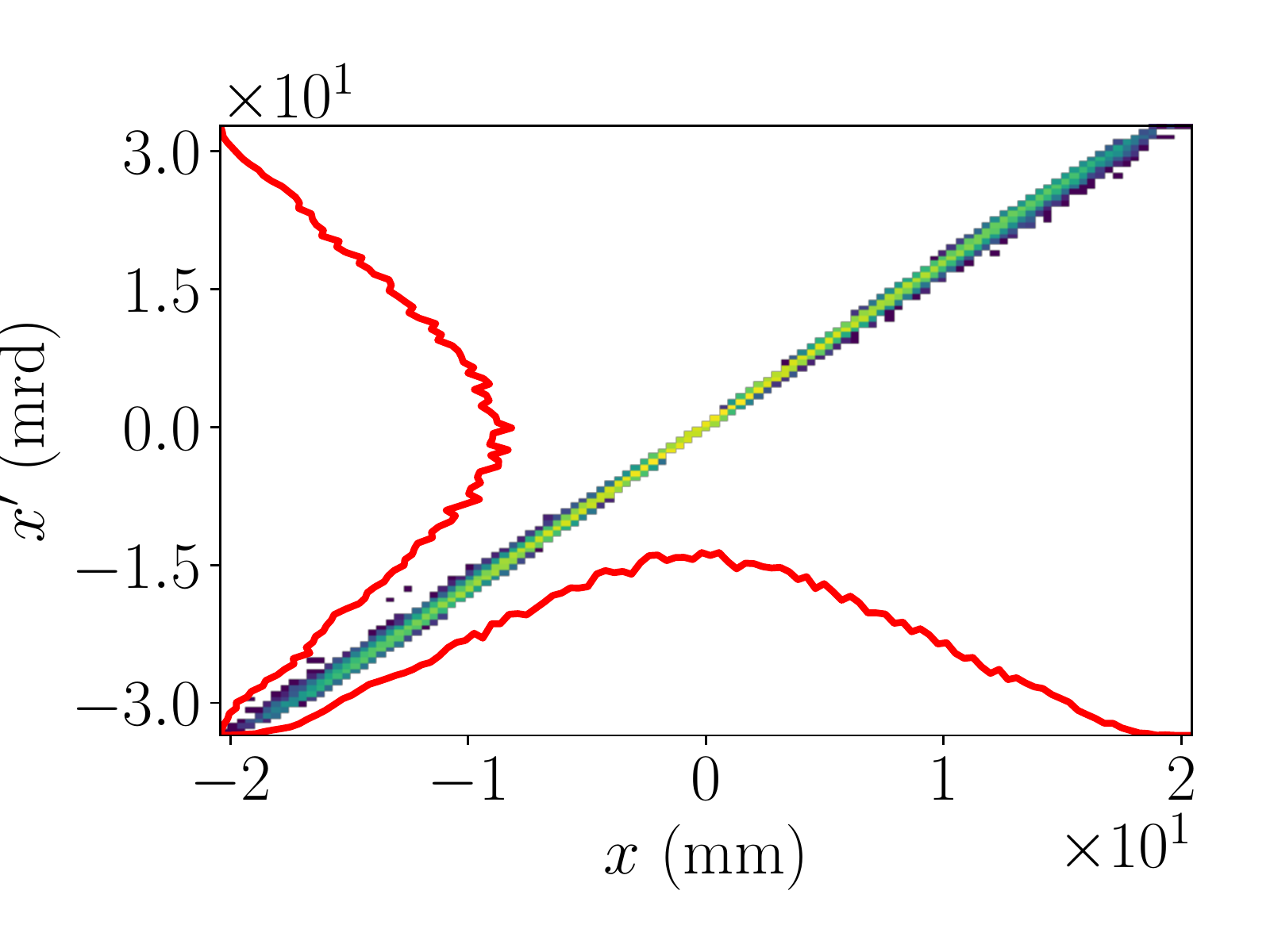}
\includegraphics[width=43mm]{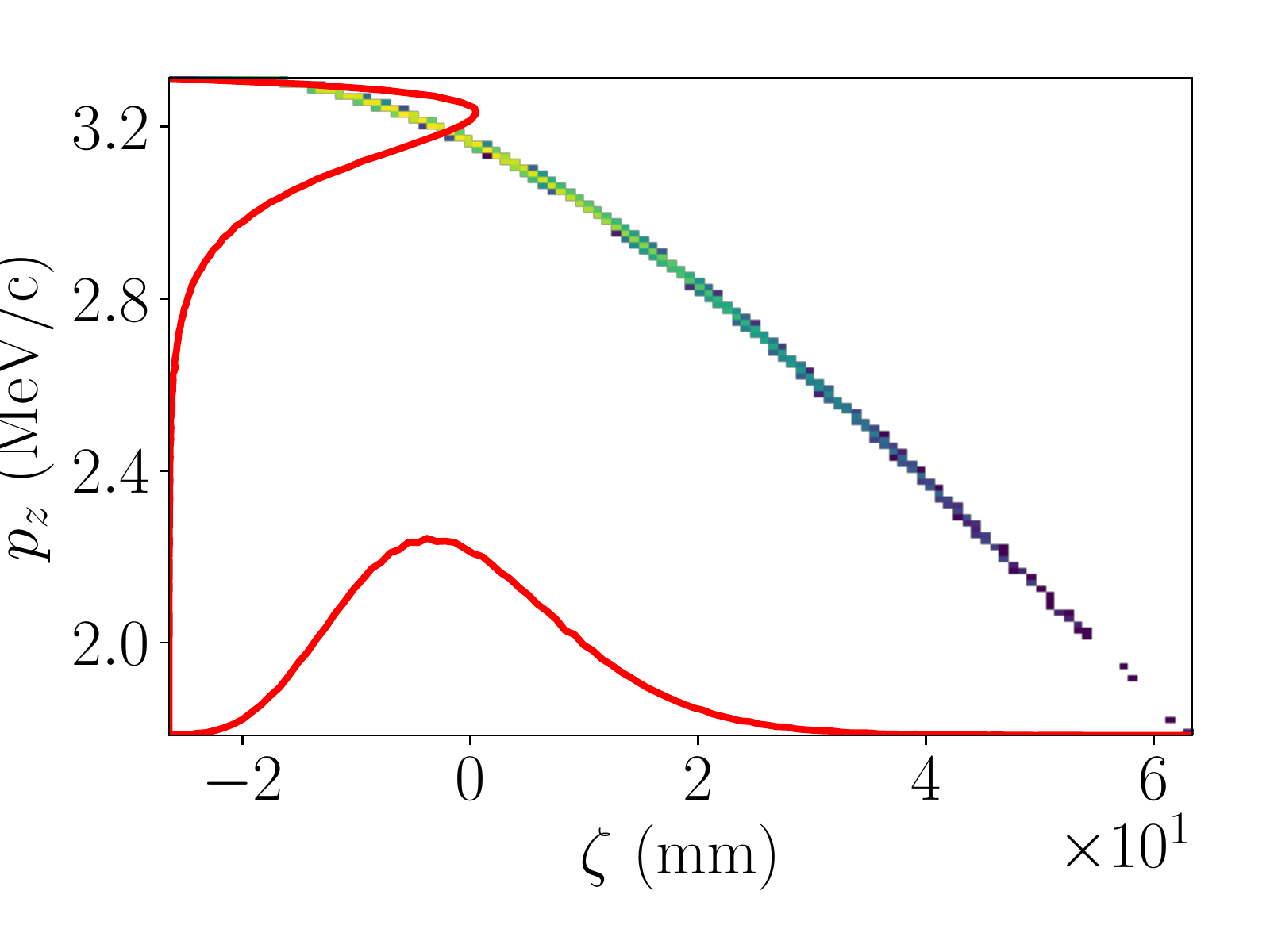}
\caption{$z =543$~mm }\label{fig:DF}
\end{subfigure}
\caption{Simulated transverse trace-space (left)  and longitudinal phase-space (right) distributions at three axial locations $z$ along the accelerator beamline ($z$ is referenced w.r.t. the cathode surface). The false color density represents the number of macroparticles (i.e. the charge) while the traces are the corresponding projections along each axis. }
\label{fig:density}
\end{figure}

A sequence of snapshots of the longitudinal phase spaces $(\zeta,\delta)$ and transverse trace spaces $(x,x')$  simulated at three different axial locations are displayed in Fig.~\ref{fig:density}. The longitudinal phase space presents a significant quadratic correlation from the early stage (starting from the exit of the bimodal cavity) while the transverse trace space displays large correlated divergences which decrease as the beam is accelerated in the SRF gun.

\section{Summary}
In summary the exploratory simulations presented in this paper suggests that the use of a field-emission source in the proposed SRF-gun prototype could be a viable option as far as the beam dynamics is concerned. Additionally, our investigations shows that some of the crucial parameter, e.g., the electron-bunch duration, have converged to their final values downstream of the 1/2-cell thereby suggesting that the 1+1/2 demo source will provide valuable information that could guide the design of the foreseen MW-class SRF gun based on a  4+1/2-cell SRF cavity. Further improvements of the modeling should include the precise optimization of the bimodal cavity to produce the required field on the cathode (e.g. to possibly accommodate cathode materials with lower enhancement factors), along with the optimization of the field profile in the SRF cavity. Nevertheless, our study provides some impetus to further investigate the technological aspects associated to the inclusion of a field-emission source and especially identify cathode materials compatible with operation in the SRF cavity. Possible contenders, beside the CNT cathodes discussed earlier, include niobium nano-ribbons~\cite{xiao} and nanocrystalline planar cathodes~\cite{euclid}.

\section{Acknowledgments}
This work was supported by the US Department of Energy (DOE) under contract DE-SC0018367 with  Northern Illinois University.  Fermilab is managed by the Fermi Research Alliance, LLC for the U.S. Department of Energy Office of Science Contract number DE-AC02-07CH11359.

\section*{References}
\bibliography{NIMA_FE_01052018.bib}
\end{document}